\documentclass{article}

\PassOptionsToPackage{numbers, compress}{natbib}

\usepackage[utf8]{inputenc} 
\usepackage[T1]{fontenc}    
\usepackage{hyperref}       
\usepackage{xurl}            
\usepackage{booktabs}       
\usepackage{amsfonts}       
\usepackage{nicefrac}       
\usepackage{microtype}      
\usepackage{xcolor}         
\usepackage{todonotes}
\usepackage{numprint}
\usepackage{multirow}
\usepackage{subcaption}
\npthousandsep{,}
\usepackage{nth}
\usepackage{tikz}
\usepackage{natbib}
\usepackage{float}
\usepackage{hyphenat}

\title{An Imbalanced Dataset with Multiple Feature Representations for Studying Quality Control of Next-Generation Sequencing}

\author{%
\small
Philipp Röchner$^{1, 2}$ \thanks{These authors contributed equally.}  \quad Clarissa Krämer$^{1*}$  \quad Johannes U Mayer$^{1, 3, 5}$ \\
\quad \small Franz Rothlauf$^1$ \quad Steffen Albrecht$^{1, 4}$ \thanks{These authors contributed equally.} \quad Maximilian Sprang$^{1, 3, 5 \dagger}$ \\
\small $^1$ Johannes Gutenberg University Mainz \\
\small $^2$ University of Southern Denmark \\
\small $^3$ Department of Dermatology, University Medical Center of the Johannes Gutenberg University \\
\small $^4$ University of Auckland \\
\small $^5$ Institute of Quantitative and Computational Biology, Johannes Gutenberg University Mainz}

\begin{document}

\maketitle

\begingroup
\renewcommand\thefootnote{}\footnotetext{%
Correspondence to: \texttt{roechner@uni-mainz.de}.}
\endgroup

\setcounter{footnote}{0}

\begin{abstract}
    \noindent
    Next-generation sequencing~(NGS) is a key technique for studying the DNA and RNA of organisms. 
    However, identifying quality problems in NGS data across different experimental settings remains challenging. 
    To develop automated quality-control tools, researchers require datasets with features that capture the characteristics of quality problems. 
    Existing NGS repositories, however, offer only a limited number of quality-related features. 
    To address this gap, we propose a dataset derived from~\numprint{37491} NGS samples with two types of quality-related feature representations. 
    The first type consists of~$34$ features derived from \textbf{q}uality \textbf{c}ontrol tools~(\texttt{QC-34} features). 
    The second type has a variable number of features ranging from eight to \numprint{1183}. 
    These features were derived from read counts in problematic genomic regions identified by the ENCODE \textbf{b}lock\textbf{l}ist~(\texttt{BL} features).\footnote{Instead of the original term used by~\citet{amemiya2019encode}, we use the terms blocklist and blocklisted regions.} 
    All features describe the same human and mouse samples from five genomic assays, allowing direct comparison of feature representations. 
    The proposed dataset includes a binary quality label, derived from automated quality control and domain experts. 
    Among all samples, $3.2\%$ are of low quality.
    Supervised machine learning algorithms accurately predicted quality labels from the features, confirming the relevance of the provided feature representations.
    The proposed feature representations enable researchers to study how different feature types~(\texttt{QC-34} vs. \texttt{BL} features) and granularities~(varying number of \texttt{BL} features) affect the detection of quality problems.
\end{abstract}

Keywords: Genomics; next-generation sequencing; quality control; bioinformatics; data quality; machine learning; benchmarking; imbalanced data; feature representations; ENCODE

\section{Background \& Summary}

Genome sequencing has deepened our understanding of biology.
In particular, next-generation sequencing~(NGS) methods read DNA and RNA snippets from biological samples in parallel, substantially reducing the time and cost of generating large amounts of biological data~\citep{satam2023next}. 
This allows, for example, clinical researchers to use NGS data to identify biomarkers for diagnosis and monitoring~\citep{menard2021clinical}. 

NGS experiments of low quality, however, can yield unreliable and difficult-to-reproduce results. 
Common quality issues include too few reads, insufficient genome coverage, and too many reads that cannot be aligned with the reference genome~\citep{taub2010overcoming}. 
Non-aligned reads can result from sample contamination, such as the presence of DNA or RNA from other sources.

To improve the quality of NGS data, several community consortia have developed quality standards for NGS experiments~\citep{sprang2024overlooked,ENCODE2011,seqc2014comprehensive,wilkinson2016andra,DBLP:journals/bioinformatics/HarrowDSRLB21}. 
The large volume of data generated by NGS experiments, however, makes manual verification difficult.
To automatically assess the quality of NGS data, machine learning approaches can be used~\citep{albrecht_seqqscorer_2021,DBLP:journals/ploscb/LiJZHCFS19}. 
Classifiers are, for example, able to automatically detect quality problems in NGS data~\cite{albrecht_seqqscorer_2021}. 

Automated detection of quality problems in NGS data with classifiers typically requires deriving the relationship between quality-related features and quality labels. 
While research consortia, such as the Encyclopedia of DNA Elements~(ENCODE)~\citep{ENCODE2011,DBLP:journals/nar/LuoHGHKLMSJLBGL20,encode2012integrated} or Cistrome~\citep{DBLP:journals/nar/ZhengWMQWSCBZML19}, provide quality labels and some quality-related features, they do not provide tabular datasets with pre-computed features suitable for developing machine learning models to detect quality problems.

To support research on automated quality control of NGS data, we introduce a dataset derived from NGS experiments with two types of quality-related feature representations: 
The first type of feature representation consists of~$34$ features~(denoted as \texttt{QC-34}), which we derived from quality control and bioinformatics tools, as described by~\citet{albrecht_seqqscorer_2021}. 
The second type of feature representation~(denoted as \texttt{BL}) captures the number of reads mapped to quality-related genomic regions included in the ENCODE blocklist~\citep{amemiya2019encode}. 
The ENCODE blocklist defines species-specific sets of quality-related genomic regions.  
While the \texttt{QC-34} features include aggregated measures of such reads, the \texttt{BL} features provide detailed information per genomic region. 
The \texttt{BL} feature representations differ by the number of considered blocklisted regions in a genome, allowing us to control the number of \texttt{BL} features between eight and \numprint{1183}. 
As the number of considered regions increases, the features provide more information on data quality. 
Both feature representations are tabular and describe the quality of the same~\numprint{37491} human and mouse samples, but capture different aspects of sample quality. 
Based on automated quality control and manual review by domain experts,~$3.2\%$ of the samples were classified as low quality and labeled as \emph{revoked}; the remaining samples were high quality and labeled as \emph{released}.

The dataset could support research in several directions. 
First, by using different types of feature representations~(\texttt{QC-34} versus \texttt{BL}) for the same dataset, researchers can study how the detection of quality problems differs between them. 
Second, by varying the number \texttt{BL} features, researchers can examine how quality control depends on the number of features. 
For example, although additional \texttt{BL} features provide more information about quality, they also increase the dimensionality of the feature space. 
This can make it harder to identify relevant patterns and may cause approaches to suffer from the curse of dimensionality: data become increasingly sparse, and distances between points become increasingly similar in high-dimensional spaces~\citep{DBLP:conf/aaai/GhosalSL24}. 

\section{Methods}

\begin{figure}[hbt!]
    \centering
    \includegraphics[width=.8\linewidth]{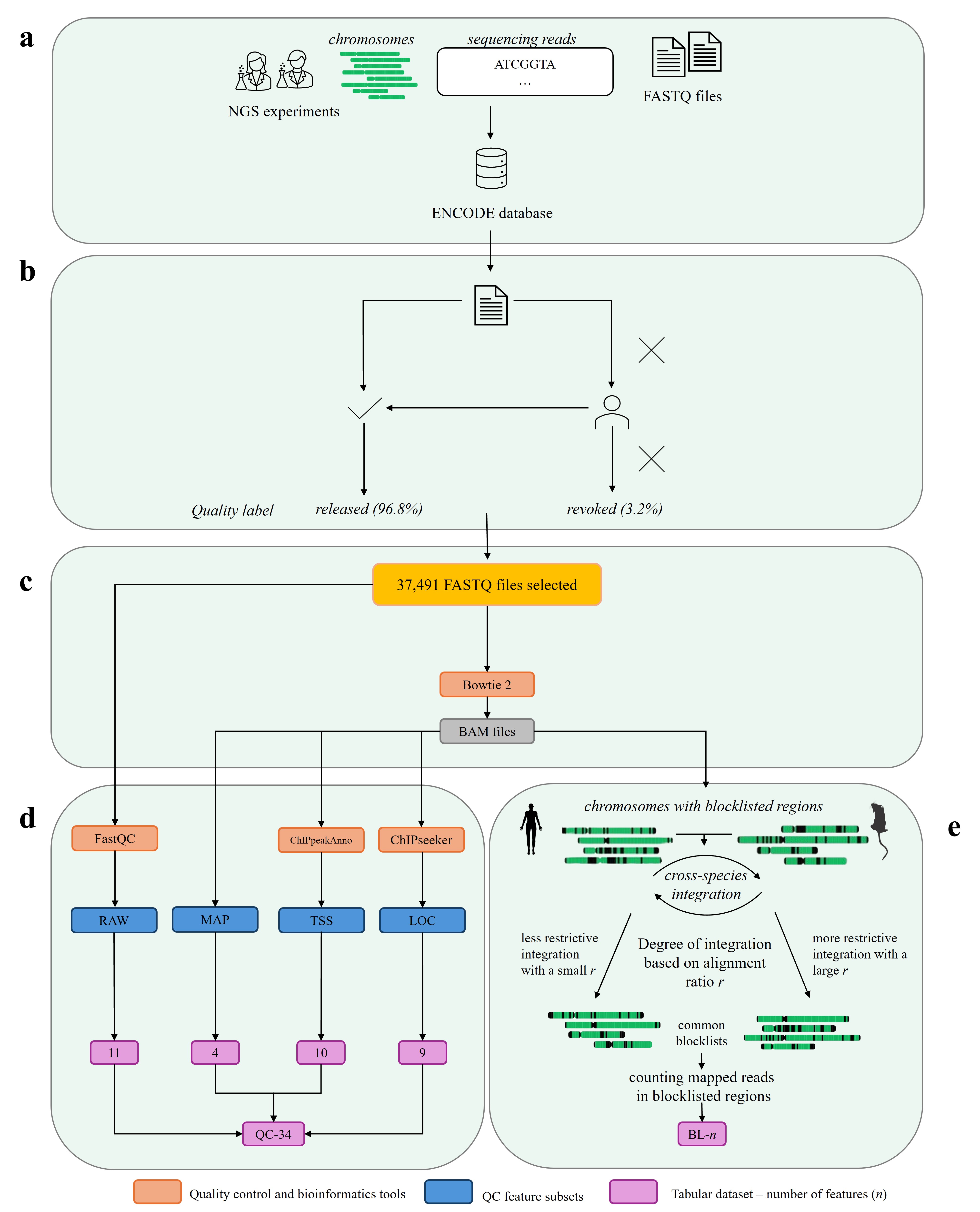}
    \caption{\textbf{Feature generation} 
    \textbf{(a)}~Researchers upload their experimental data as FASTQ files to the ENCODE database. 
    \textbf{(b)}~The experimental data is automatically reviewed based on quality metrics. 
    If these metrics indicate insufficient quality, ENCODE quality experts manually review the samples and label the reported data as \emph{released} or \emph{revoked}.
    \textbf{(c)}~We downloaded~$37,491$ FASTQ files and their associated metadata from ENCODE, then mapped the reads to the reference genomes using Bowtie 2. 
    \textbf{(d)}~The~$34$ \texttt{QC-34} features are generated using quality control and bioinformatics tools. 
    \textbf{(e)}~For the \texttt{BL} features, we first integrate the quality-related and species-specific blocklisted regions to create a combined human-mouse blocklist.
    The integration can be done with varying degrees of restriction, based on the alignment ratio \emph{r}, yielding~$n$ blocklisted regions.
    The feature values for the \texttt{BL-$n$} features are the number of reads in each of the~$n$ blocklisted regions.}
    \label{fg:fs_creation}
\end{figure}

\subsection{NGS and Quality Control Terminology}

\subsubsection{Assay Types}

NGS methods, called functional genomics assays, provide insights into gene function and regulation.
Commonly used methods are RNA sequencing~(RNA-Seq)~\citep{wang2009rna}, Chromatin Immunoprecipitation sequencing~(ChIP-Seq)~\citep{park2009chip}, DNase sequencing~(DNase-Seq)~\citep{song2010dnase}, and enhanced CrossLinking and ImmunoPrecipitation followed by high-throughput sequencing~(eCLIP)~\citep{van2016robust}. 
RNA-Seq captures gene expression, ChIP-Seq identifies where specific proteins interact with DNA, and DNase-seq detects open chromatin regions, which are parts of the chromatin accessible to cellular processes, in contrast to tightly packed~(compact and structured) DNA.
The eCLIP assay measures protein binding to RNA in the cell~\citep{van2016robust}. 

\subsubsection{File Formats} \label{sc:file_formats}

\paragraph{FASTQ} FASTQ files are the standard format for storing high-throughput sequencing data. 
Each read is represented by four lines: a sequence identifier, the sequence fragments from a sample's DNA or RNA, a separator line~(often just a +), and a quality string. 
The quality string encodes the Phred quality score for each base.
Phred quality scores reflect the confidence of the base-calling algorithm that converts the raw sequencing signal into nucleotides, which are the building blocks of DNA and RNA~\citep{cock2010sanger}. 
For a DNA or RNA read, the Phred quality score quantifies the error probability of a base call at a given nucleotide; a high score corresponds to low error probability~\citep{ewing1998base}. 

\paragraph{BAM and SAM} Binary Alignment Map~(BAM) files store read alignments to a reference genome in a space-efficient format that supports fast retrieval and analysis. 
They are the binary compressed versions of Sequence Alignment Map~(SAM) files. 
BAM files are a standard intermediate format used in sequencing workflows with SAMtools or deepTools. 
They are widely used for downstream processing, including duplicate marking, visualization in genome browsers, and read quantification in genomic regions~\citep{DBLP:journals/bioinformatics/LiHWFRHMAD09}.

\subsubsection{Quality Control and Bioinformatics Tools} 
\label{sc:feature_generation_software}

\paragraph{FastQC} FastQC is a quality control tool for high-throughput sequencing data. 
It computes summary statistics and generates visualizations to help assess the quality of raw reads. 
For example, FastQC aggregates Phred quality scores to provide an average quality score for all reads~\citep{andrews2010fastqc}.

\paragraph{Bowtie 2} Bowtie 2 is a fast, memory-efficient aligner for short-read sequencing data. 
It aligns reads, typically from FASTQ files, to a reference genome and returns the alignments in SAM or BAM format~\citep{langmead2012fast}.
Alignment software such as Bowtie 2~\citep{langmead2012fast} is also used to assess sample quality.
When mapping reads from samples to their corresponding locations in the reference genome, the resulting mapping statistics provide information on sample quality~\citep{sprang_statistical_2021}. 
For instance, the number of unmapped reads can indicate sequencing errors because incorrectly sequenced reads cannot be aligned to their genomic regions~\citep{sprang_statistical_2021}.

\paragraph{ChIPseeker} ChIPseeker is an R/Bioconductor package to annotate and visualize ChIP-Seq data. 
It accepts aligned and peak-calling data in formats such as BED or narrowPeak. 
The package maps this data to features that describe biologically meaningful elements within a genome, such as promoters, exons, or intergenic regions~\citep{DBLP:journals/bioinformatics/YuWH15}. 

\paragraph{ChIPpeakAnno} ChIPpeakAnno is an R/Bioconductor package used to annotate ChIP-Seq data. 
It facilitates overlap analysis, peak set comparison, and visualization. 
Like ChIPseeker, it works with peak files and requires that read alignment and peak calling have been performed~\citep{DBLP:journals/bmcbi/ZhuGLPLLG10}.

\subsubsection{The ENCODE Blocklist}

The ENCODE blocklist identifies quality-related regions in the genomes of several species, including those of humans and mice.
These regions are anomalous, unstructured, or highly repetitive. 
This results in high-signal regions with high read mapping rates or regions with low mappability~\citep{amemiya2019encode}. 
\citet{amemiya2019encode} derived the blocklisted regions from data collected by the ENCODE project. 

The ENCODE blocklist can be used to assess and improve the quality of NGS data. For example, ENCODE uses the fraction of reads mapped to blocklisted regions as a quality metric. 
For downstream analysis, reads mapped to the ENCODE blocklisted regions can be excluded~\citep{stark2011diffbind,zhang2008model}. 
\citet{albrecht2025seqQext} used the ENCODE blocklist to create features for automated quality control using machine learning models. 
We use blocklisted regions to generate the quality-related \texttt{BL} features~(see Section~\ref{sc:feature_generation}). 

\subsection{Data Collection}

\subsubsection{ENCODE}

The ENCODE database\footnote{\url{https://www.encodeproject.org/}} was originally a pilot project to collect information on~$1\%$ of the human genome~\citep{ENCODE2004}. 
Today, ENCODE helps scientific groups to identify all functional elements of the human genome and to make this information available to the scientific community~\citep{kagda2023datanavigationencodeportal,DBLP:journals/nar/LuoHGHKLMSJLBGL20,Jou2019portal}.
Researchers worldwide submit their experimental data to ENCODE~\citep{DBLP:journals/nar/SloanCDMSHGNHLR16}, which collects, analyzes, and publishes the data in the ENCODE database.
The database is accessible via a web portal~\citep{ENCODE2011}.

\subsubsection{ENCODE's Quality Control and Labeling} 
\label{sc:quality_labeling}

To ensure quality control, ENCODE established the Data Coordination Center~(DCC), consisting of experts in data management, bioinformatics preprocessing, and quality control of NGS data~\citep{ENCODE2011}. 
The DCC publishes guidelines, standards, and metrics for quality assessment based on reviews of NGS experiments~\citep{the_encode_project_consortium_perspectives_2020}.

When submitting their data to ENCODE, laboratories evaluate the quality of their experiments using ENCODE's open-access guidelines and standards~\citep{ENCODE2004}.  
Laboratories focusing on ChIP-Seq data, for example, must provide at least two replicates per sample. Each laboratory must also provide relevant experimental metadata. 

Based on the uploaded data, the DCC performs a two-level, semi-automated quality assessment to identify problematic experiments or samples~\citep{ENCODE2011}. 
First, ENCODE automatically identifies potentially low-quality samples by applying fixed thresholds to quality metrics, such as read length, read depth, and sequence duplication. 
Read length describes how many DNA or RNA building blocks are covered, on average, by a single read. 
Read depth quantifies the average number of times a specific DNA or RNA building block is sequenced. 
Sequence duplication, as captured by FastQC, is a measure used to detect polymerase chain reaction~(PCR) enrichment biases. 
These are three examples of measures that can be automatically generated. ENCODE considers samples that pass this automated, threshold-based quality control process to be high quality. 
These samples are labeled as \emph{released} and do not undergo further review~\citep{the_encode_project_consortium_perspectives_2020}. 

Second, DCC quality experts manually inspect samples that fail the threshold-based quality control~\citep{albrecht_seqqscorer_2021}.
After reviewing quality metrics and the broader context of the experiment, including its purpose and biological controls, DCC experts may assess samples that did not pass automated quality control as acceptable and label them as \emph{released}. 
Otherwise, samples that do not meet ENCODE standards are labeled as \emph{revoked}. 
ENCODE marks outdated samples as \emph{archived}~\citep{Jou2019portal}. 

\subsection{Sample Selection} 
\label{sc:sample_selection}

From the ENCODE database, we selected all \emph{released} and \emph{revoked} mouse and human samples for five assay types: ChIP-Seq, RNA-Seq, Poly(A)+RNA-Seq, DNase-Seq, and eCLIP.
Because RNA-Seq and Poly(A)+RNA-Seq are similar, we treated Poly(A)+RNA-Seq as RNA-Seq. 
We excluded \emph{archived} samples because their quality is unclear.

Multiple samples can belong to the same experiment. 
For paired-end experiments, we processed only the first read to avoid introducing biases into subsequent analyses, such as including reads from the same pair in both the training and test sets.
This approach has been shown to preserve relevant quality information~\citep{sprang2024overlooked}. 

We excluded samples flagged as \emph{No file available} and downloaded~$37,549$ FASTQ files, requiring~$52$ TB of disk space. 
Among these files,~$51$ were empty and therefore excluded. 
Of the remaining~$37,498$ samples, we were unable to process~$7$ FASTQ files.
We summarize the reason for exclusion in the corresponding metadata column~(see Table~\ref{tab:metadata_files}). 
The final dataset contains~$37,491$ samples.\footnote{As required by the ENCODE Data Use Policy for External Users~(\url{https://www.encodeproject.org/help/citing-encode/}), the ENCODE accession numbers of the samples used to construct the proposed dataset are included in the metadata files as described in Section~\ref{sc:data_record}.}

Table~\ref{tab:dist_metadata} shows the distribution of \emph{released} and \emph{revoked}, as well as human and mouse samples by assay type, and the overall distribution. 
The majority of samples are ChIP-Seq samples~($70.56\%$ of all samples). 
Among all samples,~$3.2\%$ are \emph{revoked}. 

\begin{table}[t]
   \scriptsize
   \centering
   \caption{Absolute~(\#) and relative~(\%) distribution of assay types in the samples. 
   The table shows the overall distribution and the distribution for \emph{released} and \emph{revoked} samples, as well as for human and mouse samples.}
   \vspace{\baselineskip}
   \begin{tabular}{l|cccc|cccc|cc}
   \toprule
   \multirow{2}{*}{Assay Type} & \multicolumn{2}{c}{Released} & \multicolumn{2}{c}{Revoked} & \multicolumn{2}{c}{Human} & \multicolumn{2}{c}{Mouse} & \multicolumn{2}{c}{Overall} \\
   \cmidrule(l){2-11}
   & \# & \% &  \# & \%  & \# & \% & \# & \%  & \# & \%  \\
   \midrule
           ChIP-Seq  & 25,491 & 70.24 & 962 & 80.17 & 23,479 & 76.18 & 2,974 & 44.57 & 26,453 & 70.56  \\
           RNA-Seq & 4,401 & 12.13 & 70 & 5.83 & 2,268 & 7.36 & 2,203 & 33.02 & 4,471 & 11.93 \\
           DNase-Seq  & 5,596 & 15.42 & 151 & 12.58 & 4,252 & 13.80 & 1,495 & 22.41 & 5,747 & 15.33  \\
           eCLIP & 803 & 2.21 & 17 & 1.42 & 820 & 2.66 & 0 & 0.00 & 820 & 2.19   \\
           \midrule
           Overall & 36,291 & 96.80 & 1,200 & 3.20 & 30,819 & 82.20 & 6,672 & 17.80 & 37,491 & 100.00  \\
       \bottomrule
   \end{tabular}
   \label{tab:dist_metadata}
\end{table}

\subsection{Feature Generation}
\label{sc:feature_generation}

Figure~\ref{fg:fs_creation} shows the generation of the proposed feature representations.

\subsubsection{\texttt{QC-34} Features}
\label{sc:qc_features}

\begin{figure}[t]
    \centering
    \includegraphics[width=\linewidth]{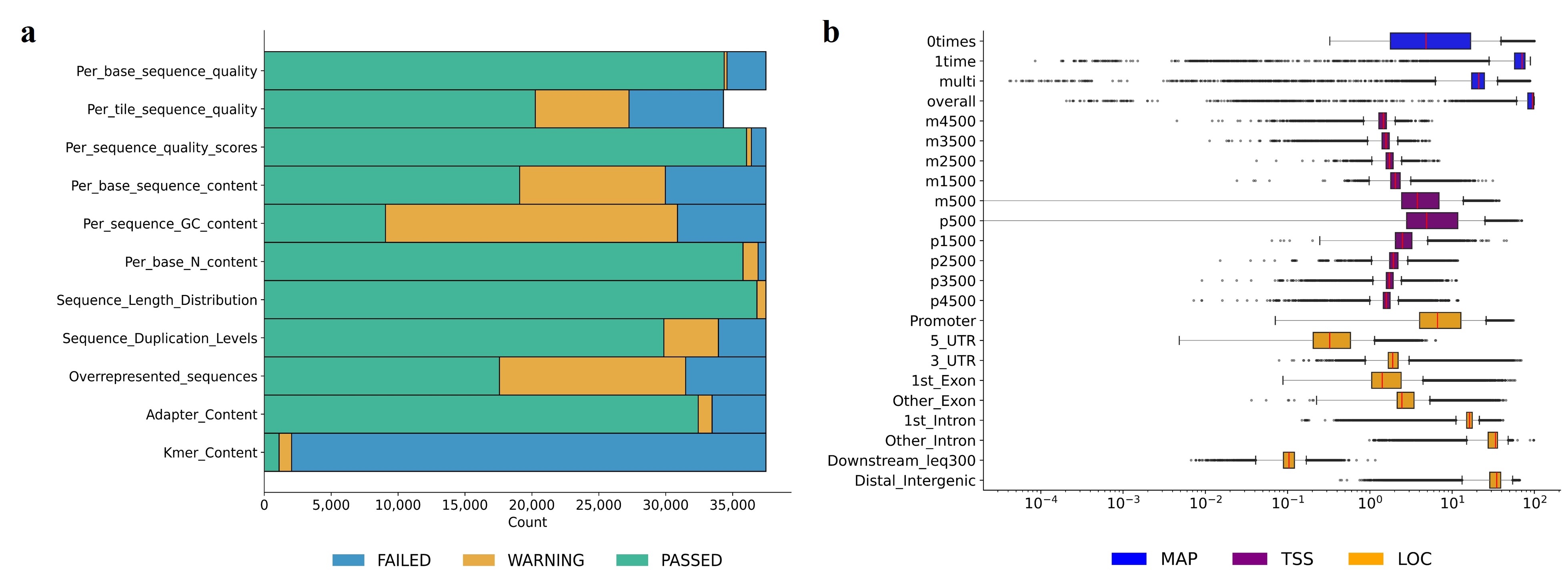}
    \caption{\textbf{Distribution of the \texttt{QC-34} feature subsets} 
    \textbf{(a)} ordinal RAW features 
    \textbf{(b)} numeric MAP, TSS, and LOC features on a logarithmic scale
    }
    \label{fg:distribution_QC_features}
\end{figure}

The \texttt{QC-34} features consist of the raw~(RAW), mapping~(MAP), transcription start site~(TSS), and location~(LOC) features, as introduced by \citet{albrecht_seqqscorer_2021}. 
In total, there are~$34$ features. 
The RAW features are ordinal with three values. 
All others features are numeric with values ranging from~$0$ to~$100$. 
Figure~\ref{fg:distribution_QC_features} lists the features and plots their distributions separately for the numeric MAP, LOC, and TSS features and the ordinal RAW features.

\paragraph{Raw Features} 

We computed the RAW features using FastQC~\citep{andrews2010fastqc} on the FASTQ files. 
FastQC assigns a \emph{FAILED}, \emph{WARNING}, or \emph{PASSED} flag based on thresholds for each metric, resulting in~$11$ ordinal features. 
The per-sequence quality score, for example, identifies subsets of a sample's sequence with low overall Phred quality scores. 
Two RAW features have missing values for some samples: the \emph{Per\_tile\_sequence\_quality} and \emph{Kmer\_Content} features, because FastQC cannot always compute these metrics.

\paragraph{Mapping Features}

MAP features are the mapping statistics generated by Bowtie 2~\citep{langmead2012fast} when mapping FASTQ files to reference genomes~(hg38, mm10).
These statistics include the percentages of reads that are mapped~(\emph{overall}), mapped multiple times~(\emph{multi}), unmapped~(\emph{0times}), or uniquely mapped~(\emph{1time}). 
High percentages of uniquely mapped reads indicate high quality, while high percentages of unmapped reads indicate low quality. 
Reads that are mapped multiple times can indicate low quality but are more context-dependent. 

\paragraph{Transcription Start Site Features}

TSS features represent the percentage of reads in~$100$ kb bins around the TSS. 
This set contains ten features, with five in each direction. 
The \emph{m4500} feature is the farthest upstream, and the \emph{p4500} feature is the farthest downstream~(see Figure~\ref{fg:distribution_QC_features}). 
Due to differences in biological context and technical factors, TSS feature values vary across samples.

\paragraph{Location Features}

LOC features describe the percentage of reads in nine functional genomic locations. 
These locations include promoter regions, enhancers, and silencers~(both covered by the feature \emph{Distal\_Intergenic}), as well as exons and introns~(gene-coding regions, divided into \emph{1st\_exon/intron} and \mbox{\emph{Other\_Exon/Intron}}), and the 3' and 5' UTR~(\emph{3\_UTR}, \emph{5\_UTR}) flanking regions~(see Figure \ref{fg:distribution_QC_features}).
Because read ratios in these regions vary with the biological context and sequencing assay, the LOC features incorporate biological and technical information. 
For example, RNA-Seq reads are distributed more evenly across the genome than peak-like ChIP-Seq assays.
For some samples, LOC features have missing values because no reads were found in the corresponding genomic regions.

To generate the LOC and TSS features, we used Bowtie 2 to provide the mapped reads in BAM format. 
We first converted the BAM files into the BED format using BEDtools~\citep{DBLP:journals/bioinformatics/QuinlanH10}, a text format required by bioinformatics tools. 
The generated BAM and BED files required~$126$ TB of storage on a high-performance computing~(HPC) file system. 
We used the Bioconductor packages ChIPpeakAnno~\citep{DBLP:journals/bmcbi/ZhuGLPLLG10} to generate the LOC features, and ChIPseeker~\citep{DBLP:journals/bioinformatics/YuWH15} to generate the TSS features. 

\subsubsection{\texttt{BL} Features} 
\label{sc:bl_features}

We derived the \texttt{BL} features from reads mapped to the reference genomes, stored as BAM files, by counting the number of reads that overlapped blocklisted regions. 

We aimed to generate a common feature representation for mouse and human samples. 
The ENCODE blocklists, however, are species-specific.
Therefore, we integrated the human and mouse blocklists to form a combined cross-species blocklist using the liftOver tool~\citep {perez2025ucsc}.
LiftOver mapped the original regions from the human blocklist to the mouse genome and the regions from the mouse blocklist to the human genome. 
To avoid one-to-many relationships, we restricted liftOver to produce a single mapped genomic region in the target genome per input region.

An important parameter for cross-species conversion is the \emph{alignment ratio} between genomes. 
This ratio specifies the minimum proportion of bases in a genomic region that must align between the two species for the region to be included in the cross-species blocklist.
This parameter determines the number of blocklisted regions remaining after conversion: a stricter alignment ratio filters out regions that differ substantially across species, retaining only those that are highly similar. 
We excluded genomic regions that had no mapped reads in any sample.

\begin{figure}[t]
    \centering
    \includegraphics[width=.7\linewidth]{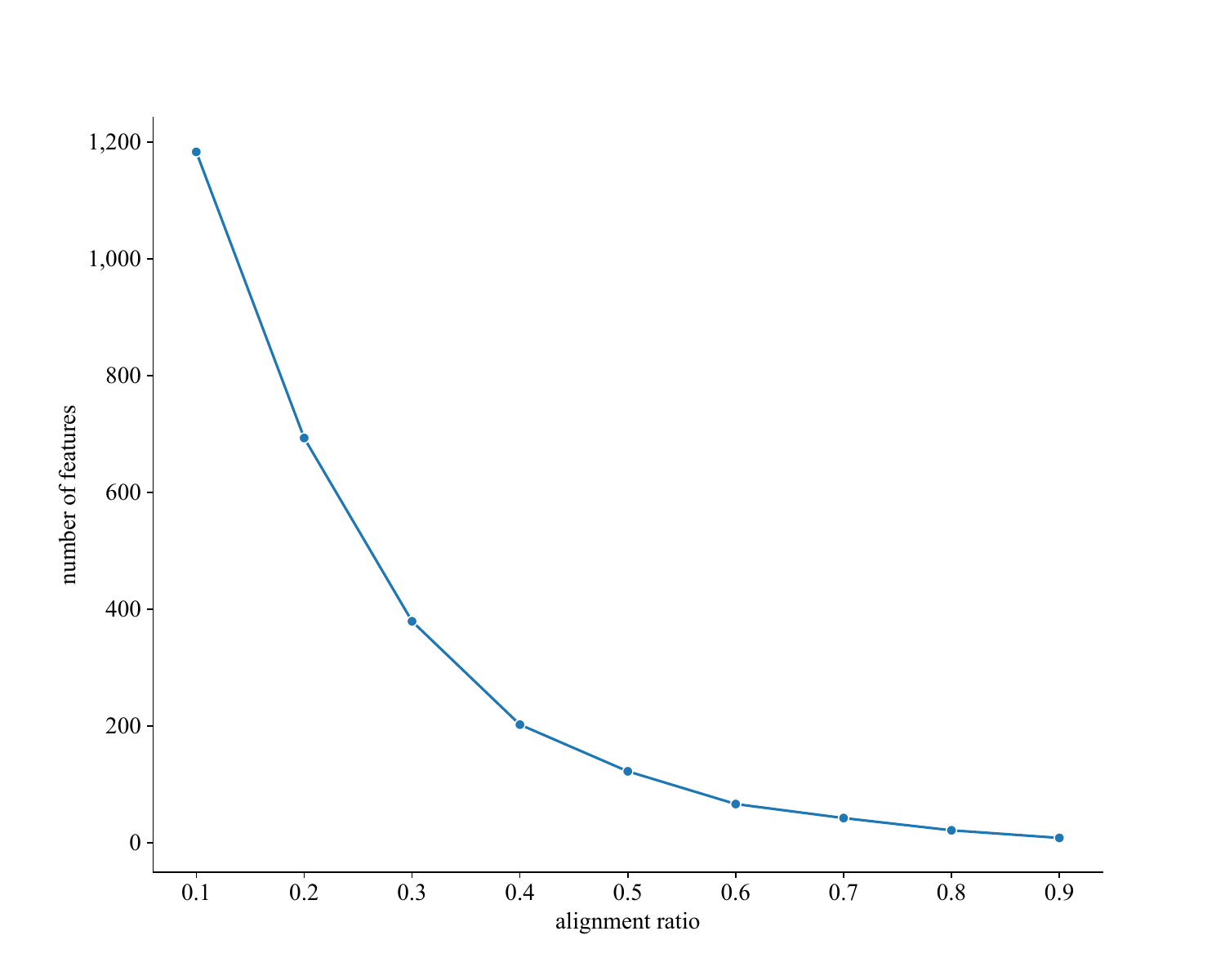}
    \caption{\textbf{Number of \texttt{BL} features depending on the alignment ratio} 
    As the alignment ratio decreases, more genomic regions with lower sequence similarity between species are included in the cross-species blocklist, resulting in more \texttt{BL} features.
    }
    \label{fg:parameter_features}
\end{figure}

Each blocklisted region corresponds to a feature, and the number of reads mapped to that region is its feature value. 
All \texttt{BL} features are numeric.
Varying the alignment ratio controls the number of \texttt{BL} features: stricter~(higher) alignment ratios yield fewer, more homogeneous features, whereas more relaxed~(lower) ratios include additional, more heterogeneous features across species. 
Importantly, \texttt{BL} features with larger alignment ratios are subsets of those with smaller ratios.
Figure~\ref{fg:parameter_features} shows how the number of features depends on the alignment ratio.

\subsection{Compute Resources} 
\label{sc:comput_resources}

\paragraph{Feature Generation}

The mapping with Bowtie 2 was the most computationally expensive step. 
Depending on the size of the FASTQ file, Bowtie 2 ran either in single-core mode or on a full 40-core compute node. 
Mapping all samples required~$1.63$ million CPU hours and an average of~$3.8$ GB of memory per sample. 
Generating the remaining features~(RAW, TSS, LOC, and blocklist) required~\numprint{18000} CPU hours and up to~$19.5$ GB of memory, depending on sample size, with an average of~$3.6$ GB per sample. 
We performed these calculations on an HPC cluster with compute nodes equipped with an Intel\textregistered{} Xeon\textregistered{} Processor E5-2630 v4, with a base frequency of~$2.20$ GHz.
The storage of all raw and mapped sequencing files occupies approximately \numprint{180} TB on the HPC file system.

\paragraph{Machine Learning Experiments}

We ran our experiments on an AMD Ryzen\texttrademark{} Threadripper\texttrademark{} 3990X 64-Core Processor with 128 GB of RAM~(architecture x86\_64) and 64 logical CPUs~(2 threads per core). 
Executing a single run of the experiments~(see Section~\ref{sc:feature_validation}) for the five assay types and proposed feature representations took approximately~$11$ CPU hours. 
Preliminary experiments required approximately three CPU hours.

\section{Data Record} 
\label{sc:data_record}

Our data is available on Zenodo.\footnote{\url{https://doi.org/10.5281/zenodo.18324916}}
The data repository contains~$15$ CSV files. 

\paragraph{\texttt{QC-34} Features}

The \emph{QC-34.csv} file contains the~$34$ quality-related features for the~\numprint{37491} NGS samples, as described in Section~\ref{sc:qc_features}, along with their quality labels~(feature \emph{status}), assay type, and organism. 
The first part of the \texttt{QC-34} feature names refers to the corresponding feature type, as described in Section~\ref{sc:qc_features}~(RAW, MAP, TSS, and LOC). 
The second part describes the quality metric.

\paragraph{\texttt{BL} Features}

The \emph{BL-$n$.csv} files contain the quality-related features for the \numprint{37491} NGS samples, as described in Section~\ref{sc:bl_features}, where~$n$ refers to the number of \texttt{BL} features. 
The files also include the samples' quality labels in the feature \emph{status}, the assay type, and the organism. 
The names of the \texttt{BL} features encode three types of information. 
The first two letters indicate whether the blocklisted region is from a human~(hs) or a mouse~(mm). 
The next two or three capital letters indicate whether the region is low mappability~(LM) or a high-signal region~(HSR) according to the ENCODE blocklist. 
The final number, separated by an underscore, refers to the genomic region in the original ENCODE blocklist. 
For example, the \emph{hsHSR\_17} feature describes the number of reads mapped to the \nth{17} blocklisted region of the ENCODE blocklist for humans.

\paragraph{Sample Metadata} 
\label{sc:sample_metadata}

The \emph{fastq\_samples\_meta.csv} file contains metadata features of the FASTQ samples derived from ENCODE. 
Table~\ref{tab:metadata_files} shows these metadata features, their missing rate, and their meaning. 
The metadata file also contains information on the~$58$ excluded files and the reason for exclusion~(see Section~\ref{sc:sample_selection}).

\begin{table}[H]
    \scriptsize
    \centering
    \caption{Metadata features of the ENCODE samples}
    \vspace{\baselineskip}
    \begin{tabular}{llp{6cm}}
    \toprule
        Feature & Missing Rate & Explanation \\
        \midrule
        Accession & $0.0\%$ & ID assigned to each experiment or sample; refers to the dataset within the ENCODE database \\
        Project & $0.0\%$ & Name or ID of the ENCODE project to which the dataset belongs; helps categorize datasets by research focus or initiative \\
        Dataset & $0.0\%$ & ID referring to the corresponding ENCODE Dataset or Experiment the sample belongs to, which contains more detailed information, see Table~\ref{tab:metadata_experiments}  \\
        Date created & $0.0\%$ & Date when the dataset was created or submitted to ENCODE; helps to track the timeline of data submissions and updates \\
        Status & $0.0\%$ & Status of the dataset, such as \emph{released} and \emph{revoked} \\
        Biosample name & $0.0\%$ & Name or ID of the biological sample used in the experiment; provides information about the tissue, cell line, or species from which the data were derived \\
        Target & $40.1\%$ & Molecular target of the experiment, such as a transcription factor or histone mark; high missing rate, as it is only relevant to some assays, such as ChIP-Seq \\
        Assay title & $0.0\%$ & Title or name of the assay or experimental technique used to generate the data, such as ChIP-Seq and RNA-Seq \\
        Batch & $46.4\%$ & ID or accession of the biosample batch the sequencing sample has been derived from; links to details about the biosample preparation in ENCODE; missing when only one batch has been prepared for the experiment; if missing, a single ID or accession can be retrieved from the experiment under \emph{Biosample accession} \\
        Donor & $0.0\%$ & Information about the donor or source of the biological sample; includes details about the individual or species from which the sample was obtained \\
        Biosample ontology & $0.0\%$ & Ontological terms describing the biology of the sample; contains information about the species, disease, and organ or cell type \\
        Platform & $0.0\%$ & Technological platform or instrument used to generate the data, such as Illumina, PacBio \\
        Library & $0.0\%$ & Information about the library preparation method, protocol, modifications, or treatments used for the experiment~(e.g., single-end, paired-end) \\
        Organ & $1.5\%$ & Organ or tissue from which the biosample was derived \\
        Not in Dataset Reason & $0.0\%$ & If samples are not included in the proposed datasets, the reason is given; for included samples, the value is \emph{included} \\
    \bottomrule
    \end{tabular}
    \label{tab:metadata_files}
\end{table}

\paragraph{Experiment Metadata} 
\label{sc:experiment_metadata}

The \emph{experiments\_meta.csv} file contains the metadata of the ENCODE experiments from which we took the FASTQ samples. 
Table~\ref{tab:metadata_experiments} shows the experiments' metadata features, their missing rate, and their meaning. 
For example, the \emph{Lab} features contain the laboratory that provided the data. 

\begin{table}[H]
    \scriptsize
    \centering
    \caption{Metadata features of the ENCODE experiments}
    \vspace{\baselineskip}
    \begin{tabular}{llp{6cm}}
    \toprule
        Feature & Missing Rate & Explanation \\
        \midrule
        ID & $0.0\%$ & Experiment accession, identical with the accessions in the Dataset and Experiment column of the metadata file \\
        Project & $0.0\%$ & Project accession; each project can contain multiple experiments and datasets \\
        Status & $0.0\%$ & Experiment status, such as \emph{released}, \emph{revoked}, or \emph{archived}; depends on the files of the experiment \\
        Biosample summary & $0.02\%$ & Description of the specimen biology, e.g. Homo sapiens K562 \\
        Biosample accession & $0.02\%$ & ID or accession for the biosample(s) prepared for a single or multiple batches used in this experiment \\
        Organism & $0.02\%$ & Organism investigated in the experiment \\
        Life stage & $0.02\%$ & Life stage of the specimen \\
        Biosample age & $17.5\%$ & Age of the sample(s)  \\
        Submitter Comment & $92.0\%$ & Comments of submitters; can contain quality-relevant information \\
        Date released & $0.0\%$ & Date when file was labeled \emph{released}  \\
        Revoked files & $79.2\%$ & List of \emph{revoked} files within the experiment \\
        Perturbed & $0.0\%$ & List of samples with perturbations  \\
        Controls & $41.7\%$ & Files that are controls \\
        Replicates & $0.02\%$ & Number of replicates \\
        Assay objective & $81.6\%$ & Objective of the assay, e.g., capture of expression, or TF binding \\
        Control type & $84.6\%$ & Type of control used in the control samples \\
        Pipeline error message & $99.8\%$ & Error messages from ENCODE's internal sample processing pipeline \\
        Alternate accessions & $96.5\%$ & Accessions of other databases such as NCBI's Gene Expression Omnibus (GEO) \\
        Lab & $0.0\%$ & Laboratory that provided the data  \\
        Biosample treatment & $90.4\%$ &  Treatment of samples with an active compound, if the given experiment had perturbations \\
    \bottomrule
    \end{tabular}
    \label{tab:metadata_experiments}
\end{table}

\paragraph{Donor Metadata} 

The \emph{donor\_ethnicity.csv}, \emph{donor\_sex.csv}, and \emph{donor\_life\_stage.csv} files provide information about the donors from whom the samples in our dataset were obtained. 
This information was derived from publicly available ENCODE metadata using donor identifiers~(see Table~\ref{tab:metadata_files}). 

The quality-related features and their metadata can be joined by accession and ID. 

\section{Technical Validation}
\label{sc:technical_validation}

\subsection{External Label Validation} 
\label{sc:external_label_validation}

We validate the ENCODE quality labels by comparing them with quality metrics derived by Cistrome. 
The Cistrome Project also provides NGS samples with associated quality information.
Some of the ChIP-Seq and DNase-Seq samples provided by Cistrome are also available from ENCODE. 
Since Cistrome generates quality flags independently of the ENCODE quality metrics, Cistrome quality flags can be used to externally validate ENCODE quality labels~\citep{liu_cistrome_2011, DBLP:journals/nar/ZhengWMQWSCBZML19}.
Unfortunately, the Cistrome database does not manually validate samples and provides fewer automatically generated quality flags than ENCODE~\citep{DBLP:journals/nar/ZhengWMQWSCBZML19}. 

\paragraph{Cistrome Quality Metrics}

We studied the following Cistrome quality metrics: the number of peaks with fold change above 10~(Peaks Fold Change Above 10), the fraction of reads in peaks~(FRiP), FastQC score, the union of DNase I hypersensitive sites~(DHS) overlapping with a union of DNase-Seq peaks~(Peaks Union DHS Ratio), and the PCR bottleneck coefficient~(PBC)~\citep{DBLP:journals/nar/ZhengWMQWSCBZML19}. 
We considered~$3,049$ ChiP-Seq samples, which were also included in the Cistrome database. 

\paragraph{Results}

First, we compare the median of the Cistrome quality metrics separately for the samples with \emph{revoked} and \emph{released} ENCODE labels.
The \emph{released} samples had a higher median 10-fold confidence peak~(Peaks Fold Change Above 10) than the \emph{revoked} samples~(median of~$545$ versus~$433$).
The median FRiP score is lower for \emph{revoked} samples~(median~$2.02\%$) than for \emph{released} samples~(median~$4.04\%$). 
The median FastQC score has a similar value for \emph{revoked} samples~(median~$38$) than for \emph{released} samples~(median~$37$). 
The median Peaks Union DHS Ratio is lower for \emph{revoked} samples~(median~$57.02\%$) than for \emph{released} samples~(median~$90.02\%$), and the median PBC score has a similar value for \emph{revoked} samples~(median~$98.95\%$) than for \emph{released} samples~(median~$98.70\%$). 

Next, we compare the overall distribution of Cistrome quality metrics for \emph{released} and \emph{revoked} samples using a Mann-Whitney U-test with a Holm–Bonferroni correction~\citep{mann_whitney,holm}. 
At a significance level of~$0.05$, three out of five Cistrome quality metrics differ significantly between the two groups: Peaks Fold Change Above 10~(p-value:~$0.18$), FRiP~(p-value:~$2.54e-05$), FastQC~(p-value:~$2.02e-04$), Peaks Union DHS Ratio~(p-value:~$2.29e-12$), and PBC~(p-value:~$0.18$). 

Overall, these results suggest that ENCODE labels and Cistrome quality flags are related.

This finding is consistent with that of~\citet{albrecht_seqqscorer_2021}. 
They used a machine learning model to relate ENCODE labels and Cistrome quality flags~\citep{albrecht_seqqscorer_2021}.
\citet{albrecht_seqqscorer_2021} trained a classifier on ENCODE data to predict ENCODE quality labels. 
When the trained classifier predicted labels for samples available only in the Cistrome database, there was a high correlation between the predicted probability that a sample is low-quality and the number of low-quality Cistrome flags~\citep{albrecht_seqqscorer_2021}. 

\subsection{Feature Validation}
\label{sc:feature_validation}

We validate that the proposed features are related to sequencing quality. 
Therefore, we identify low-quality \emph{revoked} samples based on their features using supervised classifiers.

\paragraph{Feature Sets}

We evaluated the \texttt{QC-34} features and nine \texttt{BL} feature sets. The \texttt{BL} feature sets were generated by using alignment ratios between~$0.1$ to~$0.9$ in~$0.1$ increments, resulting in a number of features between eight and \numprint{1183}. 
For identifying quality problems, we did not use the metadata features.

\paragraph{Preprocessing}

We scaled all features to the interval~$[0,1]$ by subtracting the smallest feature value from each value and dividing the result by the difference between the largest and the smallest feature values. 
For the \texttt{QC-34} feature set, missing values in the RAW features were imputed with the median. 
We determined the scaling parameters and median values using the training sets. 
Missing values in the LOC features were set to zero, as they indicate that no reads were found in the corresponding genomic regions.

\paragraph{Performance Evaluation}

We evaluated the performance of the machine learning algorithms using the area under the receiver operating characteristic curve~(AUC ROC)~\citep{DBLP:journals/pr/Bradley97}.
The AUC ROC ranges from zero to one, with higher values indicating better performance and~$0.5$ indicating random prediction. 
To account for randomness in some algorithms, we trained and evaluated each algorithm ten times and reported the average performance with its standard deviation.

\paragraph{Training Approach}

To ensure that all samples from a given experiment are either in the training or test set~(see Section~\ref{sc:data_independence}), we randomly split the dataset by experiment ID. 
The training set contains~$80\%$ of the experiments, and the test set contains~$20\%$.

\begin{figure}[t]
    \centering
    \includegraphics[width=\textwidth]{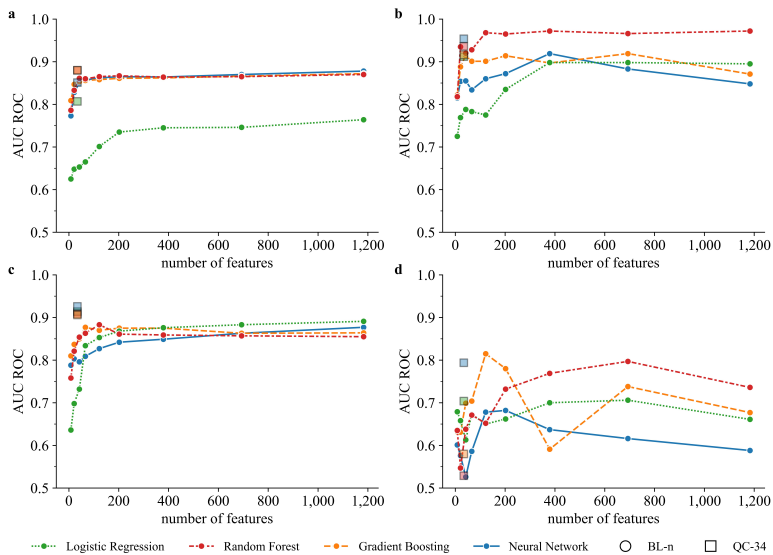}
    \caption{\textbf{Feature validation} 
    Performance of supervised classifiers for detecting \emph{revoked} samples on the test set depending on the number of \texttt{BL} features~($\circ$) and \texttt{QC-34}~($\square$) features across different genomic assays:
    \textbf{(a)}~ChIP-Seq samples,
    \textbf{(b)}~RNA-Seq samples,
    \textbf{(c)}~DNase-Seq samples, and
    \textbf{(d)}~eCLIP samples.}
    \label{fg:feat_val_su}
\end{figure}

\paragraph{Classifiers}

We used the following classifiers: Logistic Regression~(LR)~\citep{DBLP:books/lib/HastieTF09}, Random Forest~(RF)~\citep{DBLP:journals/ml/Breiman01}, Gradient Boosting~(GB)~\citep{friedman2001greedy}, and a dense Neural Network~(NN)~\citep{Goodfellow-et-al-2016}. 
For LR, RF, and GB, we used the default hyperparameters of the Python library scikit-learn~\citep{DBLP:journals/jmlr/PedregosaVGMTGBPWDVPCBPD11}. 
The NN consists of an input layer, four hidden layers with~$50$,~$20$,~$10$, and~$5$ neurons, and an output layer with a single neuron. 
We use Rectified Linear Units as activation functions for the input and hidden layers, and a sigmoid function for the output layer. 
As loss function for the output layer, we used binary cross-entropy and trained the model for~$50$ epochs with the Adam optimizer~\citep{DBLP:journals/corr/KingmaB14}.

\paragraph{Results} 

We investigated whether supervised machine learning algorithms could detect low-quality \emph{revoked} samples using the proposed features.
Figure~\ref{fg:feat_val_su} shows the AUC ROC performance of the classifiers on the test set using the \texttt{QC-34} features and depending on the number of \texttt{BL} features for the different assay types.

Except for LR, the AUC ROC values for the ChIP-Seq~(a) and DNase-Seq~(c) samples are greater than~$0.7$ for all \texttt{BL} feature sets and the \texttt{QC-34} features. 
For RNA-Seq~(b) samples, all classifiers achieved AUC ROC values above~$0.9$ for \texttt{QC-34} features and RF, GB, and NN for some \texttt{BL} feature sets.
For eCLIP samples~(d), the AUC ROC values range from approximately~$0.5$ to~$0.8$. 

Regarding the performance of the classifiers on \texttt{BL} features, RF performed as well as or better than the other classifiers for ChIP-Seq~(a), RNA-Seq~(b), and eCLIP~(d) samples for most numbers of features. 
The performance of the classifiers generally increases as the number of features increases up to approximately~$200$, except for eCLIP samples~(d). 
When there are more than~$200$ features, the performance of most classifiers stagnates for these assay types. 
In contrast, eCLIP samples~(d) generally demonstrate lower and more variable performance, with no consistent improvement as the number of features increases. 

Most classifiers performed similarly or better on \texttt{QC-34} features than on \texttt{BL} feature sets for ChIP-Seq~(a), RNA-Seq~(b), DNase-Seq samples~(c). 
For eCLIP samples~(d), the NN achieved a higher AUC-ROC on \texttt{QC-34} features than on any \texttt{BL} feature set. 
Meanwhile, RF and GB performed better on at least some \texttt{BL} feature sets than on \texttt{QC-34} features.

In general, the performance of the classifiers indicates that the proposed quality-related features accurately characterize most quality problems.

\section{Usage Notes}

\subsection{Independent Training and Test Sets}
\label{sc:data_independence}

Some samples belong to the same experiment, which can introduce dependencies if they appear in both the training and test sets.
To prevent this, we recommend two strategies:
First, all samples from each experiment are assigned exclusively to either the training or test set. 
Alternatively, one randomly selects one sample per experiment, yielding~$12,669$ independent samples that can be randomly split into training and test sets. 

Multiple samples from an experiment, however, are valuable for quantifying biological and technical variability. 
Since such applications may require working with samples from the same experiment, we provide all samples.

\subsection{Benchmarking Scenarios} 
\label{sc:benchmarking_scenarios}

The feature type, the number of features, and the sample subgroups provide three dimensions to evaluate machine learning algorithms, including their robustness and generalization. 

\paragraph{\texttt{QC-34} versus \texttt{BL} Features}

All proposed features are derived from the same $37,491$ FASTQ files, and we used BAM files as an intermediate step for all of them.
The \texttt{QC-34} and \texttt{BL} features, however, extract different quality information via separate processing pipelines, capturing complementary aspects of mapped reads from different genomic regions. 
The \texttt{QC-34} features are not summary statistics of the \texttt{BL} features; rather, they provide independent quality perspectives.
The different feature types enable researchers to compare their discriminative power.

\paragraph{Varying Number of \texttt{BL} Features}

The \texttt{BL} features for stricter~(larger) alignment ratios are subsets of those from more relaxed~(smaller) ones, as noted in Section~\ref{sc:bl_features}. 
Unlike generic statistical feature selection methods, researchers can vary the number of \texttt{BL} features based on biological properties.
\texttt{BL} features derived from a more relaxed alignment ratio yield additional blocklisted regions that are more heterogeneous between the human and mouse genomes. 
While these additional \texttt{BL} features provide more information about NGS quality than those derived from stricter alignment ratios, algorithms may struggle to detect patterns in these larger, more diverse feature sets.

\paragraph{Sample Subgroups}

The metadata provided with the dataset~(see Tables~\ref{tab:metadata_files} and \ref{tab:metadata_experiments}) enable stratified analyses, such as examining performance by assay type, species (human vs. mouse), and other biological or technical features. 
In our experiments, we observed differences in algorithm performance across assay types~(see Section~\ref{sc:feature_validation}). 
These differences reflect the underlying biological and technical complexity of quality issues, which manifest differently across assays due to variable protocols and noise sources.  
Researchers can therefore use the proposed dataset to evaluate algorithms across diverse biological scenarios and to develop robust quality control tools.

\subsection{Limitations and Future Work}
\label{sc:limitations_future_work}

\paragraph{Demographic Imbalances}

The ENCODE data used to generate the proposed dataset are not expected to be representative of the entire population, especially regarding the demographics of tissue and cell line donors. 
Although ENCODE donors are balanced by sex, certain ethnicities are underrepresented because most donors are of European ancestry. 
Training machine learning models on this dataset could unintentionally reproduce or amplify these biases. 
We therefore recommend carefully reviewing all tools developed using the proposed dataset for biases. 
For instance, models trained on the proposed dataset may not generalize well to underrepresented populations.

\paragraph{Label Quality}

Although domain experts were involved in labeling the NGS samples, the proposed dataset may still contain mislabeled samples.
We expect the \emph{released} samples to have a higher proportion of mislabeled samples than the \emph{revoked} samples because low-quality samples that pass the ENCODE quality rules incorrectly are not reviewed by domain experts and labeled as \emph{released}~(see Section~\ref{sc:quality_labeling}).

\paragraph{Other Assay Types}

Some assay types, such as single-cell RNA-Seq, differ fundamentally from those currently included in the proposed dataset. 
We therefore plan to release separate quality-related feature sets for these assay types.

\paragraph{Dataset Updates}

To capture updates to ENCODE labels and metadata, we plan to periodically provide these updates, along with quality-related features, to the community.

\section{Data Availability}

The proposed dataset, along with its feature representations and metadata files, has been deposited in Zenodo and is available at the following URL: \url{https://doi.org/10.5281/zenodo.18324916}

\section{Code Availability}

We provide a code repository that contains Python and R scripts for generating the proposed feature representations.\footnote{\url{https://github.com/Muedi/QSD/}} 
These scripts include a Python pipeline that processes a folder of FASTQ files to generate the \texttt{QC-34} features.  
Given an alignment ratio, another pipeline generates \texttt{BL} feature sets of varying sizes from the mappings produced by the first pipeline.

The code repository also contains Python scripts to reproduce the experiments in Section~\ref{sc:technical_validation}. 

\section{Acknowledgments}

We thank the ENCODE Consortium and the ENCODE production laboratories for generating and providing the data used in our study.

Parts of this research were conducted using the supercomputer MOGON 2 and/or advisory services offered by Johannes Gutenberg University Mainz~(hpc.uni-mainz.de), which is a member of the AHRP~(Alliance for High Performance Computing in Rhineland Palatinate,  www.ahrp.info) and the Gauss Alliance e.V.
The authors gratefully acknowledge the computing time granted on the supercomputer MOGON 2 at Johannes Gutenberg University Mainz (hpc.uni-mainz.de).

\section{Funding}

M.S. was supported by funding from the Einstein Early Career Researcher Award 2025, the Rise up! program of the Boehringer Ingelheim Foundation~(BIS), the ReALity Initiative of the Johannes Gutenberg University Mainz, and the Forschungsinitiative des Landes Rheinland-Pfalz. 

\section{Competing Interests}

We do not have competing interests.

\bibliographystyle{unsrtnat} 
\bibliography{01_references}

\end{document}